\documentclass[twoside,twocolumn,9pt]{article}
\usepackage{extsizes}
\usepackage[super,sort&compress,comma]{natbib} 
\usepackage[version=3]{mhchem}
\usepackage[left=1.5cm, right=1.5cm, top=1.785cm, bottom=2.0cm]{geometry}
\usepackage{balance}
\usepackage{times,mathptmx}
\usepackage{sectsty}
\usepackage{graphicx} 
\usepackage{lastpage}
\usepackage[format=plain,justification=justified,singlelinecheck=false,font={stretch=1.125,small,sf},labelfont=bf,labelsep=space]{caption}
\usepackage{float}
\usepackage{fancyhdr}
\usepackage{fnpos}
\usepackage[english]{babel}
\addto{\captionsenglish}{%
  
}
\usepackage{array}
\usepackage{graphicx}
\usepackage{dcolumn}
\usepackage{bm}
\usepackage{epstopdf}
\usepackage{epsfig}
\usepackage{droidsans}
\usepackage{charter}
\usepackage[T1]{fontenc}
\usepackage[usenames,dvipsnames]{xcolor}
\usepackage{setspace}
\usepackage[compact]{titlesec}
\usepackage{hyperref}

\definecolor{cream}{RGB}{222,217,201}

\begin{document}

\pagestyle{fancy}
\thispagestyle{plain}
\fancypagestyle{plain}{

\fancyhead[C]{\includegraphics[width=18.5cm]{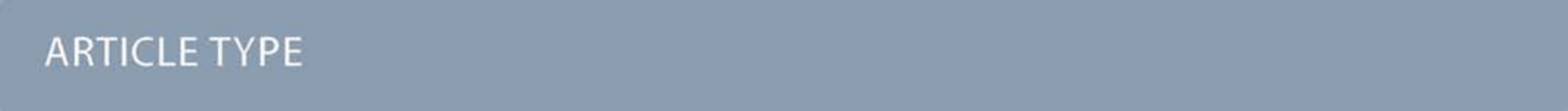}}
\fancyhead[L]{\hspace{0cm}\vspace{1.5cm}\includegraphics[height=30pt]{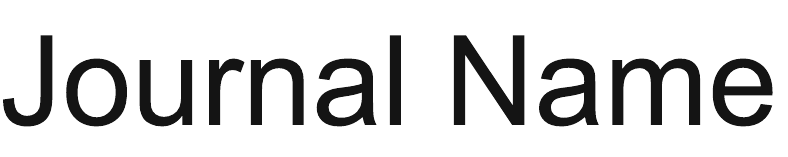}}
\fancyhead[R]{\hspace{0cm}\vspace{1.7cm}\includegraphics[height=55pt]{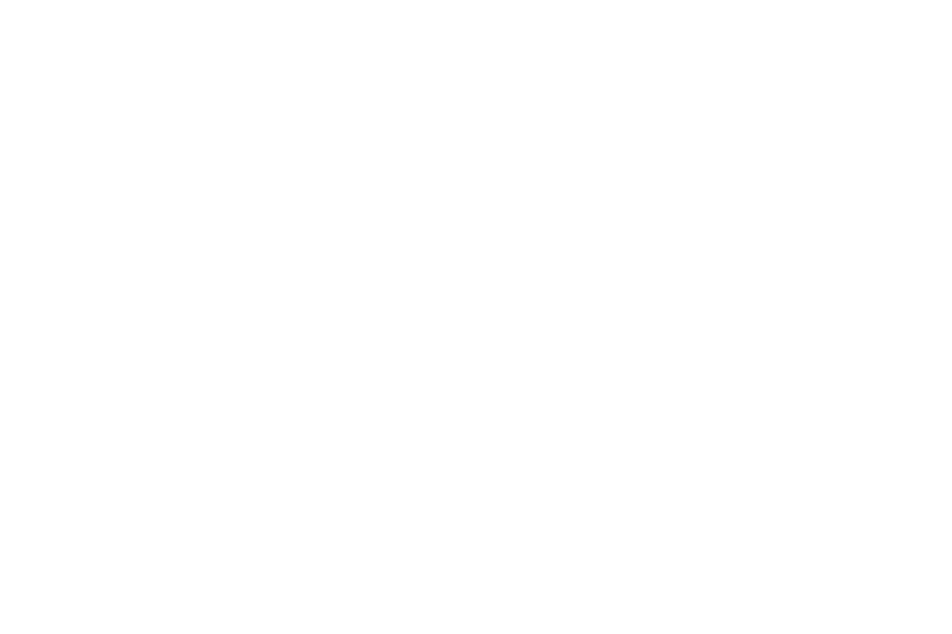}}
\renewcommand{\headrulewidth}{0pt}
}

\makeFNbottom
\makeatletter
\renewcommand\LARGE{\@setfontsize\LARGE{15pt}{17}}
\renewcommand\Large{\@setfontsize\Large{12pt}{14}}
\renewcommand\large{\@setfontsize\large{10pt}{12}}
\renewcommand\footnotesize{\@setfontsize\footnotesize{7pt}{10}}
\renewcommand\scriptsize{\@setfontsize\scriptsize{7pt}{7}}
\makeatother

\renewcommand{\thefootnote}{\fnsymbol{footnote}}
\renewcommand\footnoterule{\vspace*{1pt}%
\color{cream}\hrule width 3.5in height 0.4pt \color{black} \vspace*{5pt}} 
\setcounter{secnumdepth}{5}

\makeatletter 
\renewcommand\@biblabel[1]{#1}            
\renewcommand\@makefntext[1]%
{\noindent\makebox[0pt][r]{\@thefnmark\,}#1}
\makeatother 
\renewcommand{\figurename}{\small{Fig.}~}
\sectionfont{\sffamily\Large}
\subsectionfont{\normalsize}
\subsubsectionfont{\bf}
\setstretch{1.125} 
\setlength{\skip\footins}{0.8cm}
\setlength{\footnotesep}{0.25cm}
\setlength{\jot}{10pt}
\titlespacing*{\section}{0pt}{4pt}{4pt}
\titlespacing*{\subsection}{0pt}{15pt}{1pt}

\fancyfoot{}
\fancyfoot[LO,RE]{\vspace{-7.1pt}\includegraphics[height=9pt]{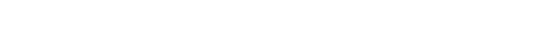}}
\fancyfoot[CO]{\vspace{-7.1pt}\hspace{13.2cm}\includegraphics{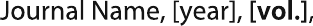}}
\fancyfoot[CE]{\vspace{-7.2pt}\hspace{-14.2cm}\includegraphics{RF}}
\fancyfoot[RO]{\footnotesize{\sffamily{1--\pageref{LastPage} ~\textbar  \hspace{2pt}\thepage}}}
\fancyfoot[LE]{\footnotesize{\sffamily{\thepage~\textbar\hspace{3.45cm} 1--\pageref{LastPage}}}}
\fancyhead{}
\renewcommand{\headrulewidth}{0pt} 
\renewcommand{\footrulewidth}{0pt}
\setlength{\arrayrulewidth}{1pt}
\setlength{\columnsep}{6.5mm}
\setlength\bibsep{1pt}

\makeatletter 
\newlength{\figrulesep} 
\setlength{\figrulesep}{0.5\textfloatsep} 

\newcommand{\topfigrule}{\vspace*{-1pt}%
\noindent{\color{cream}\rule[-\figrulesep]{\columnwidth}{1.5pt}} }

\newcommand{\botfigrule}{\vspace*{-2pt}%
\noindent{\color{cream}\rule[\figrulesep]{\columnwidth}{1.5pt}} }

\newcommand{\dblfigrule}{\vspace*{-1pt}%
\noindent{\color{cream}\rule[-\figrulesep]{\textwidth}{1.5pt}} }

\makeatother

\twocolumn[
  \begin{@twocolumnfalse}
\vspace{3cm}
\sffamily
\begin{tabular}{m{4.5cm} p{13.5cm} }

\includegraphics{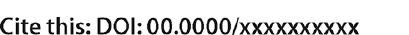} & \noindent\LARGE{\textbf{Discrete fluidization  of dense monodisperse emulsions in neutral wetting microchannels$^\dag$}} \\
 & \vspace{0.3cm} \\

 & \noindent\large{Linlin Fei,\textit{$^{a}$} Andrea Scagliarini,\textit{$^{b}$}  Kai H. Luo,$^{\ast}$\textit{$^{c}$} and Sauro Succi\textit{$^{d,b,e}$}} \\

\includegraphics{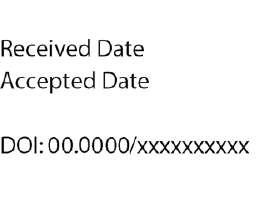} & \noindent\normalsize{The rheology of pressure-driven flows of two-dimensional dense
	monodisperse emulsions
	in neutral wetting microchannels is investigated by means of mesoscopic lattice
	simulations, capable of handling large collections of droplets, in the order of several hundreds.
	The simulations reveal that the fluidization of the emulsion 
	proceeds through a sequence of discrete steps, characterized by yielding events
	whereby layers of droplets start rolling over each other, thus leading to
	sudden drops of the relative effective viscosity.
	It is shown that such discrete fluidization is robust against loss of confinement, namely
	it persists also in the regime of small ratios of the droplet diameter over
	the microchannel width.
	We also develop a simple phenomenological model which predicts a linear relation
	between the relative effective viscosity of the emulsion and the product of the confinement
	parameter (global size of the device over droplet radius) and the viscosity ratio
	between the disperse and continuous phases.
	The model shows excellent agreement with the numerical simulations. The present work offers new insights to enable the  design of microfluidic scaffolds for tissue engineering applications and paves the way to detailed rheological studies of soft-glassy materials in complex geometries.}\\

\end{tabular}

 \end{@twocolumnfalse} \vspace{0.6cm}

  ]

\renewcommand*\rmdefault{bch}\normalfont\upshape
\rmfamily
\section*{}
\vspace{-1cm}


\footnotetext{\textit{$^{a}$Center for Combustion Energy; Key Laboratory for Thermal Science and Power Engineering of Ministry of Education, Department of Energy and Power Engineering, Tsinghua University, Beijing 100084, China. E-mail: fll15@mails.tsinghua.edu.cn}}
\footnotetext{\textit{$^{b}$Istituto per le Applicazioni del Calcolo, Consiglio Nazionale delle Ricerche, Via dei Taurini 19, 00185 Rome, Italy. E-mail: andrea.scagliarini@cnr.it}}
\footnotetext{\textit{$^{c}$Department of Mechanical Engineering, University College London, Torrington Place, London WC1E 7JE, UK. E-mail: K.Luo@ucl.ac.uk}}
\footnotetext{\textit{$^{d}$Center for Life Nano Science at La Sapienza, Istituto Italiano di Tecnologia, 295 Viale Regina Elena, I-00161 Roma, Italy. E-mail: sauro.succi@gmail.com}}
\footnotetext{\textit{$^{e}$Harvard Institute for Applied Computational Science, Cambridge, Massachusetts 02138, USA}}

\footnotetext{\dag~Electronic Supplementary Information (ESI) available: [Movies S1-S4]. See DOI: 00.0000/00000000.}



\sffamily{}

\rmfamily 

\section{Introduction}\label{sec.1}

Dense emulsions and suspensions of soft particles are widespread in natural as well as engineering applications, 
ranging from biofluids and food to pharmaceutical and cosmetic industrial processes \cite{larson1999structure,coussot2005rheometry}.
A deeper understanding of the basic rheology of these soft glassy materials (SGMs) 
is therefore beneficial to the advancement of many fields of science and technology
\cite{sollich1997rheology,bonn2017yield}.

The flow characteristics of these systems depend, often 
in a not trivial way, on many parameters, such as volume fraction, interfacial properties, deformability of the elementary,  constituents, confinement, etc.
Consequently, depending on the load conditions, SGMs display a variety
of non-Newtonian features like yielding, shear thinning/thickening, rheopexy and thixotropy
that make their rheology an outstanding open issue in non-equilibrium statistical physics \cite{larson1999structure}. During the past decades, 
significant attention has been paid to the microstructures of colloidal crystals and foams and their transformations under shear stress \cite{chen1992structural,brooks1999interfacial,debregeas2001deformation,cicuta2003shearing,dollet2015two}.
 For example, Chen \textit{et al.} studied structural changes and orientational order for dense colloidal suspensions under shear \cite{chen1992structural}. At low shear, the lattice structure will change from a crystalline state to a polycrystalline state, while at large shear, they reported the appearance of a sliding layer flow \cite{chen1992structural}. Debregeas \textit{et al.} investigated the deformation and flow of 2D foam (composed of jammed bubbles) under continuous shear in a Couette geometry \cite{debregeas2001deformation}. They found rapid decay of the average velocity from the moving inner wall to the fixed outer wall, as well as large velocity fluctuations with self-similar dynamical structures inbetween. They also developed a stochastic model relating the plastic flow to the stress fluctuations.  Cicuta \textit{et al.} reported rheological measurements on dense soft glass with an interfacial stress rheometer (ISR) \cite{cicuta2003shearing}. They
found an interesting response of the elastic modulus\cite{cicuta2003shearing}:
a transition from viscous liquid towards an elastic solid with the increase of concentration, and an inverse transition (from elastic to viscous) with the increase of shear frequency. 

The shear rheology of a wide class of SGMs \cite{Coussot,Hoehler,Becu} 
can be described by the Herschel-Bulkley relation \cite{HB} between the applied stress $\sigma$ and the responsive shear rate $\dot{\gamma}$,
\begin{equation}\label{eq:HB}
\sigma = \sigma_Y + K \dot{\gamma}^c
\end{equation}
In Eq. (\ref{eq:HB}), $\sigma_Y$  
is the yield stress, below which the material does not flow, instead it reacts elastically (showing solid-like behaviour) 
to the imposed load; for $\sigma > \sigma_Y$ the material flows, in general with a non-Newtonian constitutive law characterized by the two 
phenomenological parameters $K$ (the consistency factor) and $c$ (the flow index) that depend on the detailed 
microstructure of the material \cite{coussot2005rheometry}. For $c<1$ the material is said to be {\it shear-thinning},
whereas  $c>1$ corresponds to {\it shear-thickening}. For dense suspensions of solid 
particles the phenomenology can be even richer: a number of rheometric studies, 
provided evidence of flow curve hysteresis and a {\it discontinuous} shear-thickening 
\cite{Seto,Fernandez,wyart2014discontinuous,Kawasaki} (see also \cite{Brown} and references therein). On the other hand,
discontinuous shear-thinning has remained so far more elusive. Chen and Zukoski \cite{Chen} observed, in 
{\it constant-stress rheometry} of concentrated suspensions with crystalline order, and {\it well beyond yield}, 
that a critical stress exists at which the power law index 
of the flow curve changes. Similar results have been obtained in numerical simulations of adhesive dispersions of soft disks
\cite{Irani}.
It must be stressed, though, that the occurrence of such discontinuous shear-thinning/thickening behaviours can be
assessed in shear rheometry, where the shear stress is {\it homogeneous} across the sample, and for values of the applied
load {\it above the yield stress}, but not in microchannel flows,
unlike what was recently proposed in \cite{foglino2017flow}. 

Different from the previous study within the Couette geometry \cite{chen1992structural,debregeas2001deformation,cicuta2003shearing,Chen,Irani},
	the present work focuses on the deformation and flow of two-dimensional (2D) emulsions in the Poiseuille geometry. 
When a pressure difference is imposed along a neutral wetting microchannel,
the shear stress varies linearly from wall to wall, being maximum at the sidewalls and zero in the channel center. With the increase in pressure difference, it is possible to achieve a fluid-like state near the wall, i.e., $ {\sigma _w} > {\sigma _Y} $, where $ {\sigma _w} $ indicates the wall stress.
However, if the material develops a non-vanishing yield stress $\sigma_Y$, there will always be a region, 
in the middle of the channel, where the emulsion is below yield and must be expected to move as a plug. For the meso-constituents near the wall, different from sliding on a non-wetting microchannel \cite{fei2018mesoscopic}, they are expected to stick on the neutral-wetting wall.
In such a setup, the way a soft-glassy material fluidizes, i.e. the plug-to-fluid flow transition, 
still lacks a full understanding.  As we mentioned, the microstructure in which the meso-constituents (droplets in the case of emulsions) are 
arranged is known to play a determining role; 
two structural indicators are crucial in this respect, namely the polydispersity,
that measures the statistical distribution of sizes of the meso-constituents, and 
the degree of order in their spatial arrangement (crystalline vs amorphous) \cite{SaintJalmes,ShibaOnuki,benzi2015internal}.  In addition, a high volume fraction $ \Phi $ of the dispersed droplets is necessary to support such a microstructure transition. For example, recent studies on the T1 transition in complex microchannels are mainly based on emulsions/foams with $ \Phi >\sim 0.85$ \cite{Gai,gai2019timescale,vecchiolla2019}. 

The rest of the paper is organized as follows. In Section \ref{sec.2},
we give a brief introduction of the numerical model: the two-species Lattice Boltzmann (LB) model with suitable pseudo-potential interactions. The main features of the adopted numerical model are discussed and well validated.
In Section \ref{sec.3}, we prepare  perfectly 2D monodisperse dense emulsions (constituted of soft droplets), with crystalline order, 
confined in microchannels and driven by a pressure gradient. 
The droplets are packed at the jamming volume fraction,
such that the emulsion behaves as an elastic solid for vanishingly low applied loads. 
We focus, then, on the fluidization of the material and find that the latter takes place as
a discontinuous process with the increase of pressure gradient, characterized by successive local yielding events,
where layers of droplets, starting from those closer to the boundaries,
are more and more deformed and 
roll over each other. Scanning the imposed forcing, we observe that these events cause sudden 
drops of the relative effective viscosity between from one plateau to another; the relative effective viscosity is, therefore, ``quantized'', in the sense that it takes values only over a discrete set. 
The first and largest jump
occurs for a critical pressure difference, which we find to be such that the wall
stress equals the yield stress of the material. 
We show, moreover, that the value of the relative effective viscosity in the plug state depends, 
in a way that we characterize quantitatively, on the 
viscosity ratio between the disperse and continuous phases and confinement parameter
(channel height over droplet radius). Finally, a brief summary and some research perspectives are given in Section \ref{sec.4}.

\section{Model description and validation}\label{sec.2}

 \begin{figure}[!htbp]
	\center {
		{\epsfig{file=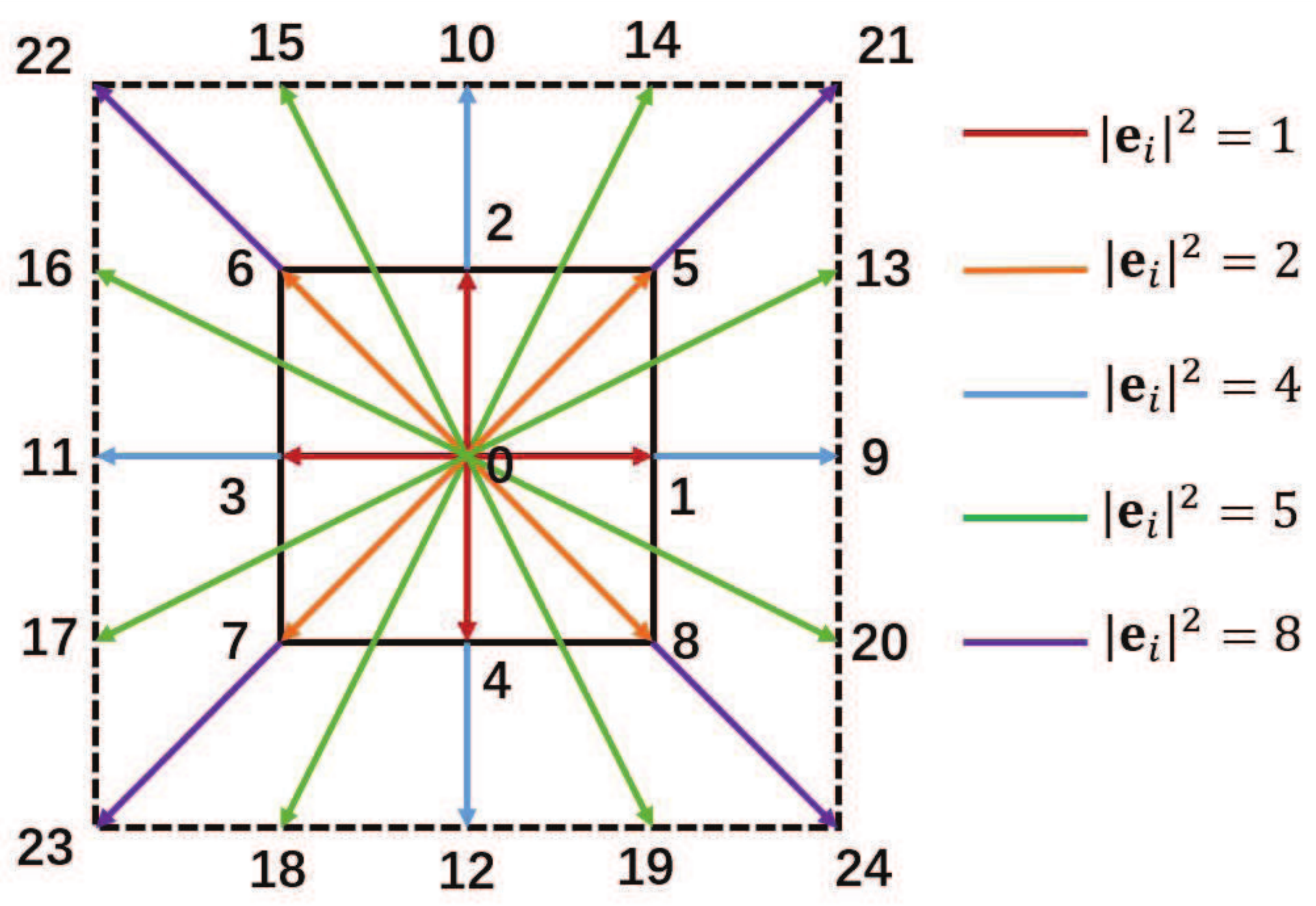,width=0.45\textwidth,clip=}}\hspace{0.0cm} \\ 
	}
	\caption{The discrete lattice used in our two-species LB model. The fluid lives in the sub-lattice (D2Q9), while the interactions extend to the full lattice (D2Q25).}
	\label{Fig-lattice}
\end{figure}
We perform numerical simulations based on a two-species lattice Boltzmann (LB) model, with 
suitable pseudo-potential interactions. Recently, the LB method has become a very efficient and powerful simulation method for complex flows \cite{li2016lattice_softmatter,fei2017consistent,colosqui2012mesoscopic,fei2018three,li2016lattice,ISI:000466701800016,ISI:000468255400008,chantelot2018water,ISI:000444957300007,ISI:000401593600016,lei2019generalized,montessori2019modeling}. 
    The LB equation takes the form 
	\cite{higuera1989lattice,higuera1989boltzmann,qian1992lattice,benzi1992lattice}:
	\begin{equation}\label{e1}
	{f_{k,i}}({\bf{x}} + {{\bf{e}}_i}\Delta t,t + \Delta t) = {f_{k,i}} - \frac{{\Delta t}}{{{\tau _k}}}({f_{k,i}} - f_{k,i}^{eq}) + \Delta t{F_{k,i}}.
	\end{equation}
	where ${f_{k,i}}$ is the probability of finding a particle of species 
	$k$ ($k = 1,2$) at the space-time point $({\bf{x}},t)$, moving along the \textit{i}th ($i = 0,1,...,8$, see Fig. \ref{Fig-lattice}) direction ${{\bf{e}}_i}$ 
	of a regular D2Q9 lattice \cite{qian1992lattice}. The right-hand side, computed at $({\bf{x}},t)$, 
	represents the time relation towards local equilibrium $f_{k,i}^{eq}$ on a time scale ${{\tau _k}}$ and ${F_{k,i}}$ is a 
	forcing term incorporating the effects of the total force $ {{\bf{F}}_k} $ 
	acting upon each species \cite{fei2018mesoscopic}. The kinematic viscosity for each component is related to the relaxation time by $ {\nu _k} = ({\tau _k} - 0.5\Delta t)c_s^2 $, where $ \Delta t = 1$ and $ {c_s} = \sqrt {1/3} $ are the time step and lattice sound speed, respectively.
	During the evolution of Eq. (\ref{e1}), the density and momentum for each species are calculated as,
\begin{equation}\label{e5}
{\rho _k} = \sum\limits_{i = 0}^8 {{f_{k,i}}} , ~~~{\rho _k}{{\bf{u}}_k} = \sum\limits_{i = 0}^8 {{f_{k,i}}} {{\bf{e}}_i} + \Delta t{{\bf{F}}_k}/2,
\end{equation}
and the total velocity of the fluid mixture is updated by $ {\bf{u}} = \sum\nolimits_k {{\rho _k}{{\bf{u}}_k}/} \rho  $,
where the total density is $\rho  = \sum\nolimits_k {{\rho _k}} $.

For the two-species system, a repulsive force  between species, promoting a positive surface tension, is give as usual\cite{shan1993lattice}, 
\begin{equation}\label{e8}
{\bf{F}}_k^r =  - {\rho _k}({\bf{x}}){\sum _{\bar k}}{G_{k\bar k}}\sum\limits_{i = 0}^8 {w({{\left| {{{\bf{e}}_i}} \right|}^2}){\rho _{\bar k}}({\bf{x}} + {{\bf{e}}_i})} {{\bf{e}}_i},
\end{equation}
where $ {G_{k\bar k}}{\rm{ = }}{G_{\bar kk}} $ is the strength coefficient for the inter-component interaction, and the weights are $ w(0) = 4/9 $, $ w(1) = 1/9 $, and $ w(2) = 1/36 $. A crucial characteristic of our adopted two species LB model is the competing mechanism between a short range attraction and a mid-range repulsion within each species \cite{benzi2009mesoscopic}. The short-range attraction acts between the nearest-neighbor D2Q9 lattice nodes, while the mid-range repulsion acts between the next-to-nearest-neighbor lattice nodes extending up to the D2Q25 lattice ${{\bf{e}}_j}$ ($j = 0,1,...,24$, see Fig. \ref{Fig-lattice}). The competing interaction is explicitly written as \cite{benzi2009mesoscopic}
\begin{equation}\label{e7}
\begin{split}
{\bf{F}}_k^c &=  - {G_{k,1}}{\psi _k}({\bf{x}})\sum\limits_{i = 0}^8 {w({{\left| {{{\bf{e}}_i}} \right|}^2}){\psi _k}({\bf{x}} + {{\bf{e}}_i})} {{\bf{e}}_i} \\
&- {G_{k,2}}{\psi _k}({\bf{x}})\sum\limits_{j = 0}^{24} {p({{\left| {{{\bf{e}}_j}} \right|}^2}){\psi _k}({\bf{x}} + {{\bf{e}}_j})} {{\bf{e}}_j},
\end{split}
\end{equation}
where $ {G_{k,1}} $ and $ {G_{k,2}} $ are introduced to tune the strengths of the two interactions, respectively, and the weights for the D2Q25 lattice are  $ p(0) = 247/420 $, $ p(1) = 4/63 $,  $ p(2) = 4/135 $,  $ p(4) = 1/180 $,  $ p(5) = 2/945 $ and  $ p(8) = 1/15120 $. The pseudo-potential function originally suggested in \cite{shan1993lattice,shan1994simulation}, $ {\psi _k}({\rho _k}) = {\rho _0}(1 - {e^{ - {\rho _k}/{\rho _0}}}) $ (with an identical reference density $ {\rho _0}=1.0 $ for each component) is adopted. Supplemented with the body force ${\bf{F}}_k^b$, the total force imposed on each species is ${{\bf{F}}_k} = {\bf{F}}_k^b + {\bf{F}}_k^r + {\bf{F}}_k^c$.

 The above definition of ${\bf{F}}_k^c$ is to mimic the spatially complex (non-monotonic) interactions among molecules within each species. For example, the interplay between the competing interactions can give rise to a rich density-field configuration, which  has proved fairly successful in reproducing many features of 
 soft flowing systems, such as structural frustration, aging, elastoplastic rheology, in confined and unbounded flows \cite{sbragaglia2012emergence,benzi2015internal,dollet2015two,scagliariniCOLSUA}.
 In addition, it has been shown that by appropriately tuning the strengths of the two interactions, a positive disjoining pressure $\varPi$, which emerges as a repulsive force per unit area between opposing interfaces, can be obtained at sufficiently low film thickness \cite{sbragaglia2012emergence}.  Moreover,  such a positive disjoining pressure can be tuned independently of the viscosity ratio between the two species, as recently proposed in  \cite{fei2018mesoscopic}. 
 
\begin{figure}[!htbp]
	\center {
		{\epsfig{file=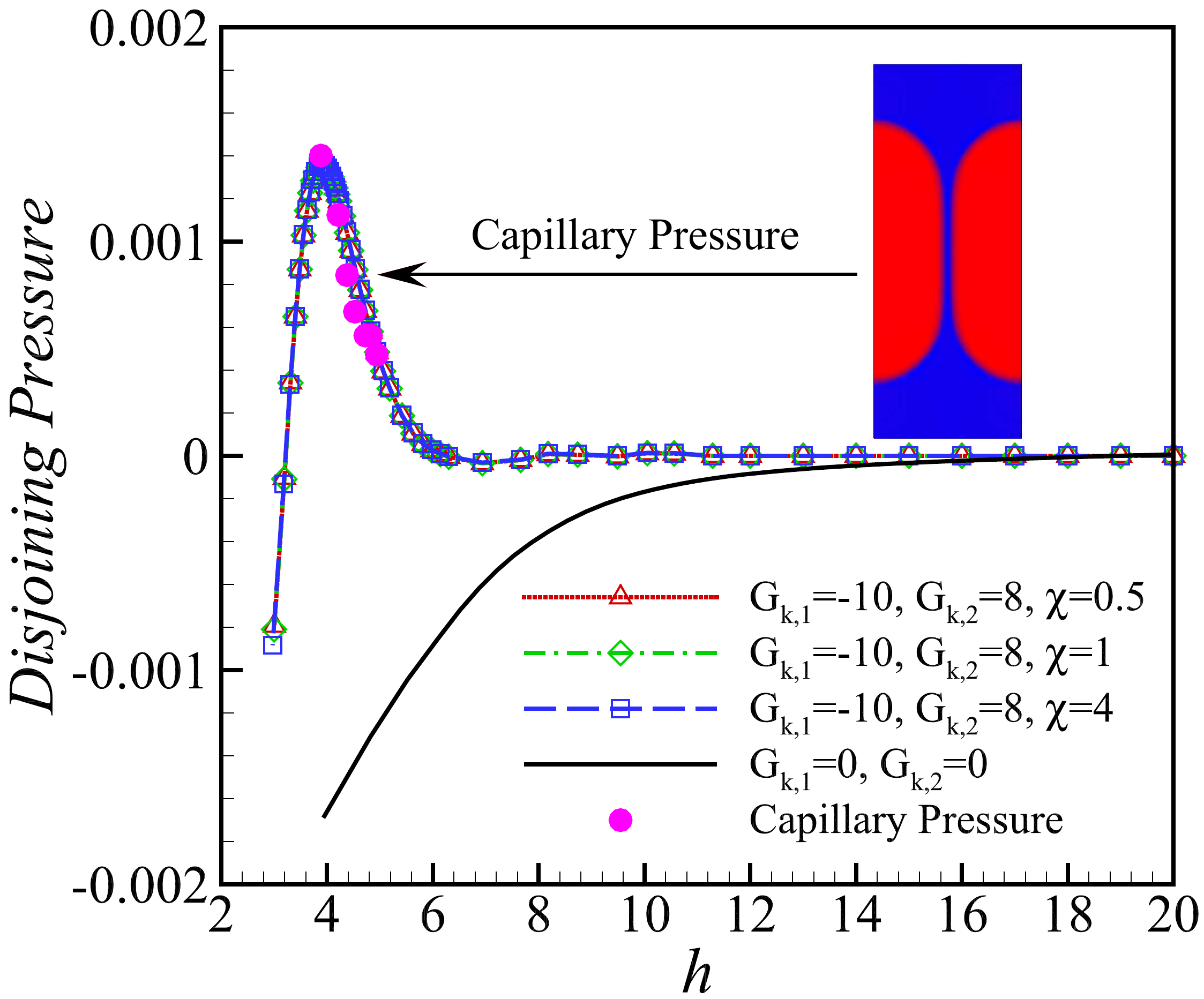,width=0.43\textwidth,clip=}}\hspace{0.0cm} \\ 
	}
	\caption{Lines: theoretical disjoining pressure curves by Eq. (\ref{edis}) for different cases; Pink filled circle: capillary pressure curve based on the two hemispherical droplets system at  $ {G_{k,1}} =-10$, $ {G_{k,2}} =8$ and $\chi=1$; Inset: a stable two
		hemispherical droplets system.  The positive disjoining pressure between the close-contact interfaces stabilizes the thin film inbetween.}
	\label{Fig-Disjoining-Pressure}
\end{figure}

To verify the adopted LB model, we first  consider two flat interfaces among three fluid domains (occupied by species 1, 2 and 1, respectively), separated by a distance \textit{h} (the width of the middle fluid domain). According to \cite{bergeron1999forces,sbragaglia2012emergence}, the disjoining pressure $ \varPi $ is defined as 
 \begin{equation}\label{edis}
  \int_{\varPi (h = \infty )}^{\varPi (h)} h d\varPi  = {\gamma _f}(h) - 2\gamma ,
 \end{equation}
where $ \gamma $ is the surface tension and $ \gamma _f $ is the overall line tension, defined as the integral of the mismatch between the normal ($ {P_{xx}} $) and tangential ($ {P_{yy}} $) components for the pressure tensor \cite{shan2008pressure}, i.e., $ {\gamma _f}(h) = \int_{ - \infty }^{+\infty}  {({P_{xx}} - {P_{yy}})}dx $. Supplemented with the boundary conditions, $ {\gamma _f}(h \to \infty ) = 2\gamma $ and $ \varPi (h \to \infty )=0 $, the disjoining pressure $ \varPi $ at a given $ h $ can be calculated using standard numerical integration methods \cite{fei2018mesoscopic}. The obtained disjoining pressure based on the above theory is shown in Fig. \ref{Fig-Disjoining-Pressure}. It can be seen that for the original pseudo-potential model \cite{shan1993lattice} ($ {G_{k,1}} =0$, $ {G_{k,2}} =0$, i.e., without the competing interaction), the disjoining pressure is always negative and decreases with decreasing $ h $. By tuning the competing interaction appropriately ($ {G_{k,1}} =-10$ and $ {G_{k,2}} =8$), the disjoining pressure first increases with the decrease of $ h $ and then goes down, achieving positive values in between. Moreover, the disjoining pressure profile is independent of the viscosity ratio $\chi$ between the two species within a moderate range. 

To test the above theory, we then preform a numerical measurement based on the method suggested in \cite{sbragaglia2012emergence}: two hemispherical droplets (species 1) are pushed against each other to form a thin film (species 2) inbetween. According to the mechanical equilibrium,  the disjoining pressure in the thin film must be equal to the capillary pressure at the curved interface, defined as the pressure difference from both sides of the curved interface. By changing the radii of the droplets, we can obtain the capillary pressure at different values of the film width $ h $. A good agreement between the theoretical disjoining pressure and the measured capillary pressure can be clearly seen in Fig. \ref{Fig-Disjoining-Pressure}. It is noted that the capillary pressure curves at varying viscosity ratio $\chi$ are essentially identical and only the case of  $\chi=1$ is shown in Fig. \ref{Fig-Disjoining-Pressure}. 

\begin{figure*}[!htbp]
	\center {
		{\epsfig{file=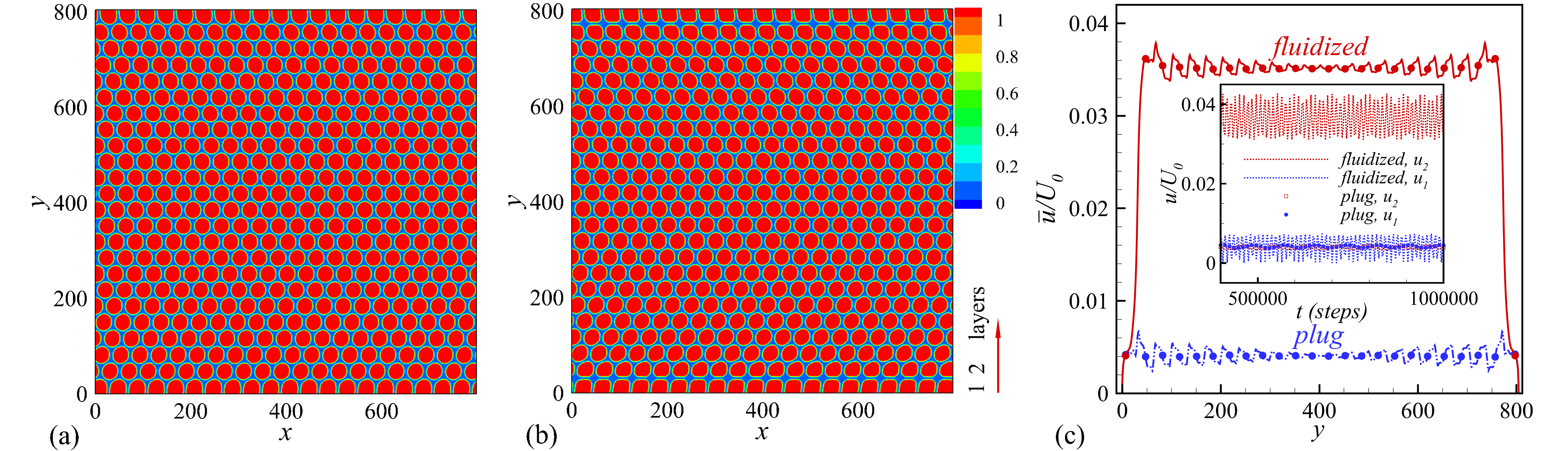,width=1\textwidth,clip=}}\hspace{0.0cm} \\ 
	}
	\caption{ (a) A snapshot of the simulated emulsion with $\xi = 44.7$ and $\chi=4$, 
		in the ``plug'' state. Red and blue colors represent high and low density of droplets.
		(b) Same as in (a) but for the ``fluidized'' state. The first two layers of droplets (near to the wall) are marked as layer $1$ and $2$, 
		respectively. (c) Velocity profiles normalized by the Poiseuille velocity $U_0$ for the same 
		pressure gradient, for the plug and fluidized states.
		Inset: velocity of the first two layers of droplets, $u_1$ and $u_2$, in the plug  and in the fluidized state .}
	\label{Fig1}
\end{figure*}
Therefore, the novelty of our improved numerical model, recently proposed in \cite{fei2018mesoscopic},
resides in a number of original features which prove key to probing the high packing
regime typical of dense emulsions, in particular: 
(i) the emergence of a positive disjoining pressure between close-contact interfaces,
which stabilizes the thin films between  
colliding droplets and prevents their coalescence,
(ii) the possibility to tune the dynamic viscosity ratio between the two species,
independently on surface tension and disjoining pressure, within a moderate
range, and (iii) the highly efficient implementation of dense emulsions with up to hundreds of dispersed droplets,  due to the explicit and simple algorithm for only two 
species probabilities ${f_{k,i}}$ ($k = 1,2$).

\section{Numerical results and discussion}\label{sec.3}

\subsection{Discrete fluidization}
We prepare the monodisperse soft suspensions by packing equal sized droplets,
of radius $R=18$ lattice units, in
a 2D microchannel, with the length and width varied in the range
$ L \in [200,800] $ and $ H \in [201,804] $, respectively;
no-slip boundary conditions for the velocity and neutral wetting conditions for the two fluids
(contact angle of ${\rm{9}}{{\rm{0}}^ \circ }$) are imposed on both walls. Large values of $L \times H$, 
with up to $\approx 500$ droplets, are necessary to verify that the discontinuous 
rheological response is not an artifact due to small sizes, and let us unravel the impact of confinement. We would like to 
stress that such sizes would be computationally unfeasible with other LB methods 
where coalescence is prevented by introducing a distinct species for each droplet \cite{foglino2017flow}, even equipped with advanced acceleration algorithm based on the exclusion principle \cite{Spencer-2011}.

We introduce the two non-dimensional parameters that play a key role in the dynamics: 
the confinement parameter $\xi=H/R$, where $R$ is the 
droplet radius (large $\xi$ means low confinement and vice versa), and the viscosity ratio $\chi=\mu_D/\mu_C$, 
where $\mu_D$ and $\mu_C$ are the dynamic viscosity of the droplets and continuous phase, respectively. The viscosity ratio is tuned by changing $\mu_D$ with fixed  $\mu_C=0.1$. The pressure difference $\Delta p$ between inlet and outlet is imposed via a homogeneous body force $\Delta p/L$; 
the lattice interaction parameters are chosen as, $ {G_{k\bar k}}{\rm{ = }}{G_{\bar kk}} =2.33$,  $ {G_{k,1}} =-10$, $ {G_{k,2}} =8$,
 to realize a surface tension ($\gamma \approx 0.02$), 
such that the corresponding Laplace pressure for droplets 
is $p_L \equiv \gamma/R \approx 10^{-3}$ (LB units) and to achieve a positive disjoining pressure, 
sufficient to prevent droplet coalescence (see Fig. \ref{Fig-Disjoining-Pressure}). 
We performed a set of runs spanning the 2D parameter space $(\xi,\chi)$,
at effective volume fraction of the 
dispersed phase $\Phi \approx 0.9$ \cite{MBW95,MBW96}. 

In Fig. \ref{Fig1} [(a) and (b)], we show two typical snapshots of the simulated system 
(the density field of the dispersed phase fluid $\rho_D$ is depicted, 
the red/blue colours indicating high/low values; see also the Supplemental Movies S1 and S2): 
an emulsion with viscosity ratio $\chi=4$ and under low confinement, $\xi=44.7$, is  
subject to low [panel (a)] and high 
forcing [panel (b)].
The corresponding streamwise velocity profiles ${\bar u(y)}$ (averaged along the $x$-direction and in time, 
over the steady state) are plotted in Fig. \ref{Fig1}c, with blue/red lines standing for low/high pressure gradients, respectively. 
In both cases the profiles flatten (as compared to the parabolic profile expected for a Poiseuille flow in a simple liquid), 
signalling that a portion of bulk
remains below yield. However, while for the lower forcing this region extends over the whole volume, 
such that the material moves coherently as a solid block (a ``plug''), for  
the higher forcing the droplets in the two layers next to the walls are more deformed and the droplets in the next-to-nearest 
layers roll over those in direct contact with the wall periodically: in other words, the emulsion starts to be ``fluidized'' (see Movie S2). 
This is confirmed in the inset of the same figure, where we plot the velocities of the first 
two layers of droplets (counting from the bottom wall), called $u_1$ and $u_2$: in the plug state these two values coincide 
(no deformation within the material, as it would be for a solid), whereas,
at increasing the pressure difference, they bifurcate, with $u_2 > u_1$ and periodic velocity oscillations.
\begin{figure}[!htbp]
	\center {
		{\epsfig{file=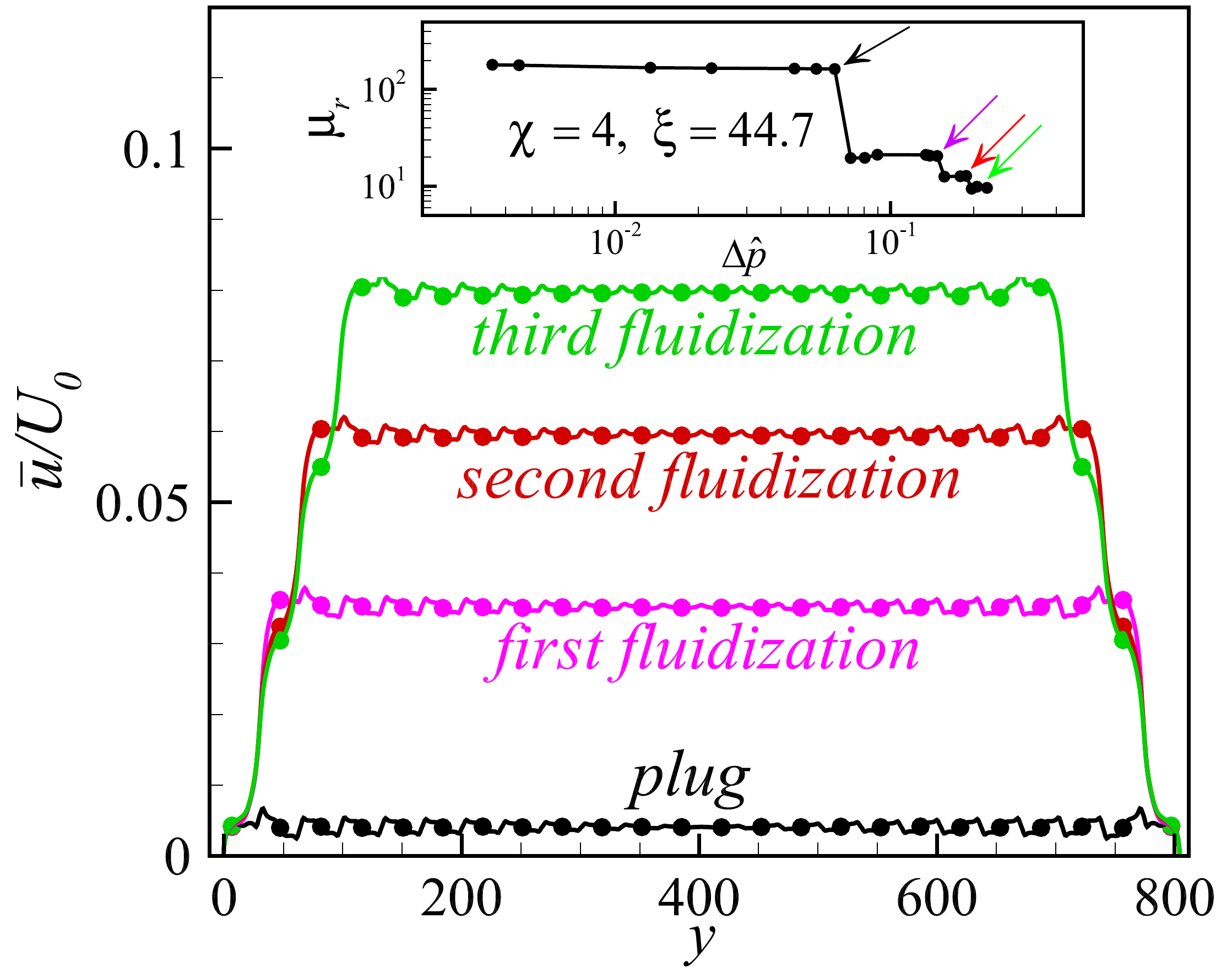,width=0.45\textwidth,clip=}}\hspace{0.0cm}  
	}
	\caption{Velocity profiles corresponding to the plug state and the first three fluidizations, for 
		a dense emulsion with $\chi=4$ and $\xi=44.7$. Inset: relative effective viscosity vs non-dimensional pressure difference;  the locations corresponding to the different fluidizations are highlighted with arrows.}
	\label{Fig2}
\end{figure}

One may wonder whether other activation events of this kind will involve
more and more droplet layers, for larger and larger imposed pressure gradients.
This is, indeed, what happens, as we can grasp from the main panel of  
Fig. \ref{Fig2}, by inspection of the velocity profiles (same system as in Fig. \ref{Fig1}, i.e.
with $\chi=4$ and $\xi=44.7$), where, by second and third fluidization, we mean the activation of
the third and fourth droplet layers (and the symmetric ones on the top wall side) that extends the region of shear
localization towards the bulk.

\begin{figure}[!htbp]
	\center {
		{\epsfig{file=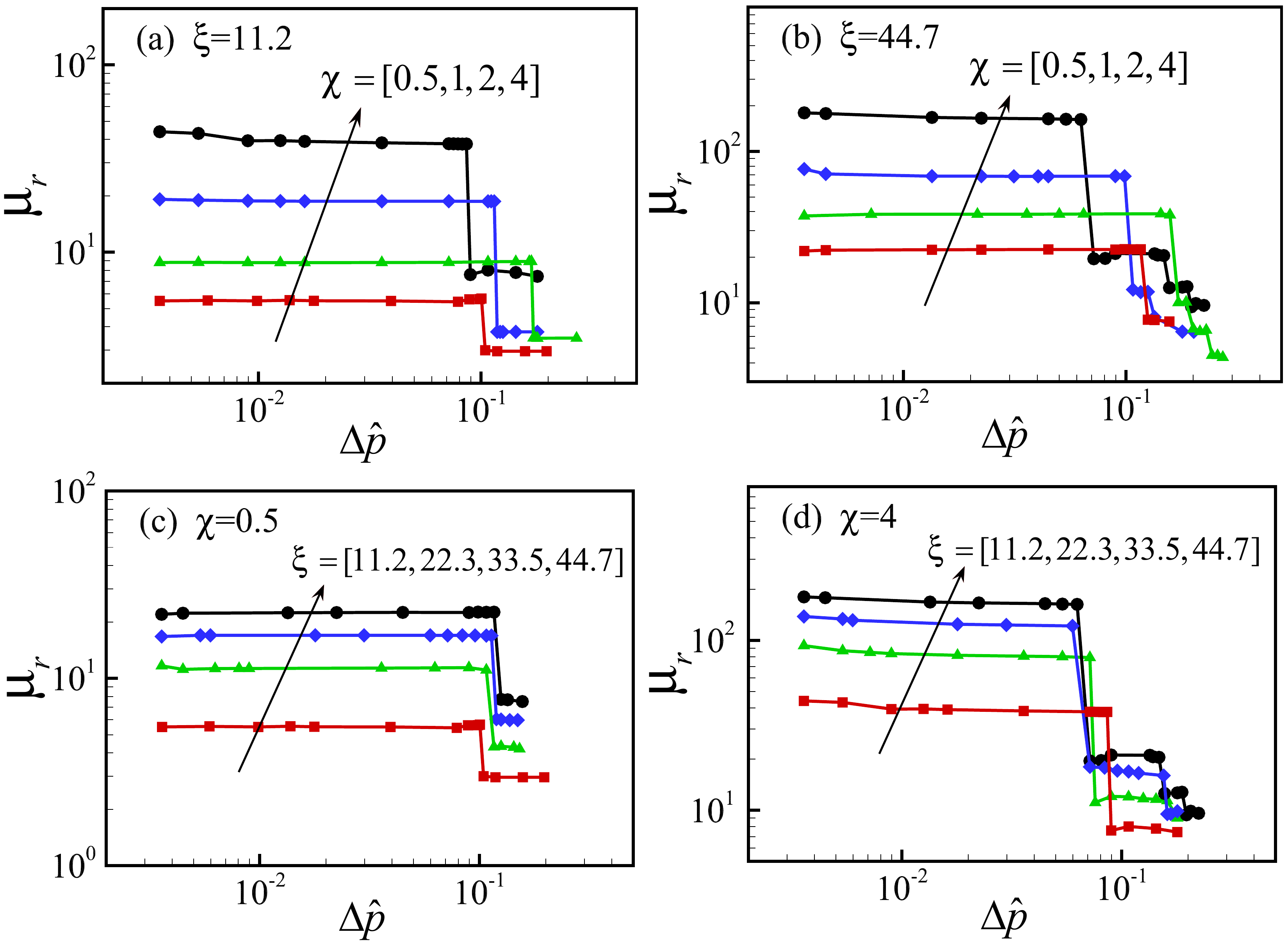,width=0.49\textwidth,clip=}}\hspace{0.0cm}  
	}
	\caption{
		Relative effective viscosity of the emulsion, $\mu_r$, as a function of
		the non-dimensional
		applied pressure difference
		$\Delta \hat{p}$. Panels (a) and (b): various viscosity ratios $\chi$ between
		dispersed and continuous phases at fixed confinement ($\xi=11.2$, panel (a); $\xi=44.7$, panel (b)).
		Panels (c) and (d): various confinement parameters $\xi$ at fixed viscosity ratio ($\chi=0.5$, panel (c);
		$\chi=4$, panel (d)).}
	\label{Fig3}
\end{figure}

To characterize the different flow regimes, at changing the imposed pressure drop, we define the 
relative effective viscosity, $ {\mu_r} = {Q_0}/Q $ \cite{ZhouPozrikidis}, where
$ {Q_0} $ is the flow rate for the pure continuous phase for a given $ \Delta \hat p $ (defined later) and 
$ Q = \int_0^H {\bar u(y)dy} $  is the emulsion flow rate. 
The dependence of  $ {\mu_r} $ on $ \Delta \hat p $ for various viscosity ratios
and confinement parameters is shown in Fig. \ref{Fig3}: for fixed
confinement and varying viscosity ratio (\ref{Fig3}(a), $\xi=11.2$, and \ref{Fig3}(b), $\xi=44.7$),
and for fixed viscosity ratio and varying confinement (\ref{Fig3}(c), $\chi=0.5$, and \ref{Fig3}(d),
$\chi=4$), respectively.
In every data set, we observe an initial steep decrease of $\mu_r$ for very low pressure differences ($\Delta \hat p \le {10^{ - 3}}$, not shown in the figure),
followed by a clear-cut step-wise behaviour: $\mu_r$ is roughly constant, $\mu_r \approx \mu_r^P$ 
(the superscript $P$ standing for ``{\it plug} state''), over a certain range of $ \Delta \hat p $ values, and then   
suddenly jumps to a lower plateau $\mu_r^F$ (``{\it fluidized} state''), 
as the pressure difference is increased  beyond a ``critical" threshold $\Delta \hat p_{c}$.
The applied pressure drops (on the $x$-axis) are made dimensionless, hereafter, as follows:
\begin{equation} \label{eq:rescdp} 
\Delta \hat{p} = \frac{\Delta p}{p_L}\left(\frac{H}{2L}\right),
\end{equation}
where the latter expression represents the ratio between the wall stress and the Laplace pressure of the droplets.
As $\Delta \hat{p}$ is increased further, the relative effective viscosity develops more additional jumps (see Figs. \ref{Fig3} (b)and (d), as well as the inset of Fig. \ref{Fig2}),
and related after-jump plateaux. 
Essentially, the material responds elastically to the 
increasing applied load (whence the $\mu_r$-plateaux), up to a point where the stress cannot 
be sustained any more and the system {\it yields}
(corresponding to the $\mu_r$-jumps), 
with successive boundary trains of droplets that start to roll over each other
(see Movies S3 and S4). This phenomenology is what we dub
{\it discrete fluidization}, with an accompanying {\it discrete} relative effective viscosity.
Let us underline that jumps at higher $\Delta \hat{p}$ are smaller and smaller,  suggesting that the fluidization process is recovering continuity and eventually the material
starts flowing as a viscous, non-Newtonian, fluid.

Recently, Foglino  \textit{et al.} also found the first and largest jump for emulsions with unit viscosity ratio in a small channel ($\chi=1$ and $\xi=12$) \cite{foglino2017flow}. In the present work, such a behaviour is further confirmed and significantly extended, and its robustness is well verified, in the sense that we find even more jumps in a larger parameter space ($\chi$, $\xi$). In addition, the first jump was interpreted as a discontinuous shear thinning behaviour \cite{foglino2017flow}. We would like to stress that the shear-thinning process happens in a material which is well beyond yield, i.e.,
$\sigma > \sigma_Y$, as we have discussed in the Introduction.
In our simulations, the first discontinuity in the $\mu_r$ vs $\Delta \hat{p}$ curve, instead, coincides for all $\xi$'s, upon the proper rescaling Eq. (\ref{eq:rescdp}),
at roughly the same critical pressure drop $\Delta \hat{p} \approx 0.1$. Remarkably, the latter 
condition is equivalent to 
\begin{equation}
\Delta p \frac{H}{2L} \equiv \sigma_w \approx 0.1 P_L = 0.1 \frac{\gamma}{R},
\end{equation}
where $\sigma_w$ is the wall stress. 
The expression at the rightmost hand side is comparable with the theoretical value,
obtained by Princen \cite{Princen83}, of the yield stress of a perfectly monodisperse 
two-dimensional, highly concentrated emulsion, at a volume fraction $\Phi \approx 0.9$. 
These considerations suggest that the fluidization transition occurs when
the imposed pressure difference is such that $ \sigma_w \sim \sigma_Y$, i.e. when the maximum imposed 
stress across the material equals its yield value, lending to the idea that the
origin of the discontinuity 
can be ascribed to local yielding events, rather than the discontinuous shear-thinning events proposed in \cite{foglino2017flow}.

\begin{figure}[!htbp]
	\center {
		\includegraphics[scale=0.35]{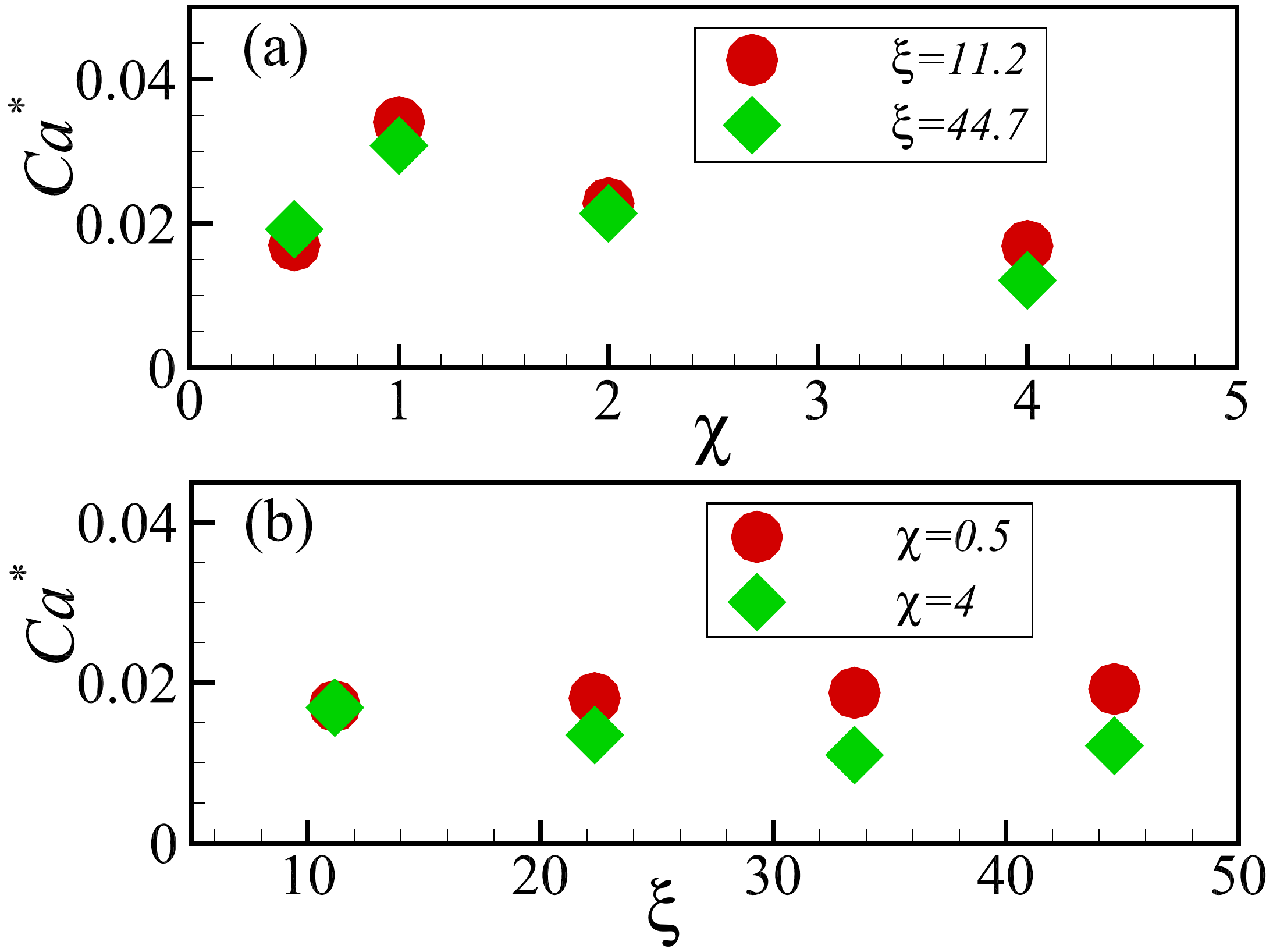}\hspace{0.0cm}  
	}
	\caption{Variation of the maximum Capillary number in the plug state, $ C{a^*} $, with  viscosity ratio  $\chi$ (panel (a)), and with confinement parameter  $\xi$ (panel (b)).}
	\label{Fig-Ca}
\end{figure}

Taking the channel height $ H $ as the characteristic length, $ V = Q/H $ as the characteristic velocity, the Reynolds number and Capillary number can be defined, respectively, as $ {Re}  = {\rho _D}VH/{\mu _D} $ and $ Ca = {\mu _D}V/\gamma $. For all the cases considered in the present work (i.e., Fig. \ref{Fig3}), the $ {Re} $ and $ Ca $ span approximately from 0.015 to 500 and 0.0006 to  0.7, respectively. It is noted that the maximum Capillary numbers in the plug state, indicated as $ C{a^*} $, for all the considered cases, almost collapse around  0.01 $ \sim $ 0.03, as shown in Fig. \ref{Fig-Ca}, while we do not find a similar collapse for critical Reynolds number. This criterion is consistent with the previous study for the dislocation dynamics in 2D emulsion crystal in a tapered channel that horizontal slip planes parallel to the $ x $ axis are observed at $ Ca > \sim 10^{ - 2} $ \cite{Gai}. In the experiments, the extrusion speed of the crystal needs to be sufficiently small ($ Ca < 10^{ - 2}$), so that the flowing foam is in the solid-like plug state (below yield) and the spatial distributions of T1 evens remain localized in distinct rearrangement zones \cite{Gai,vecchiolla2019}. In other words, the collapse of the critical Capillary number, $ C{a^*} \sim {10^{ - 2}} $, further supports our previous argument that the plug-to-fluid transitions indicate local yielding events.

\subsection{Dependence on confinement and viscosity ratio}

\begin{figure}[!htbp]
	\center {
		{\epsfig{file=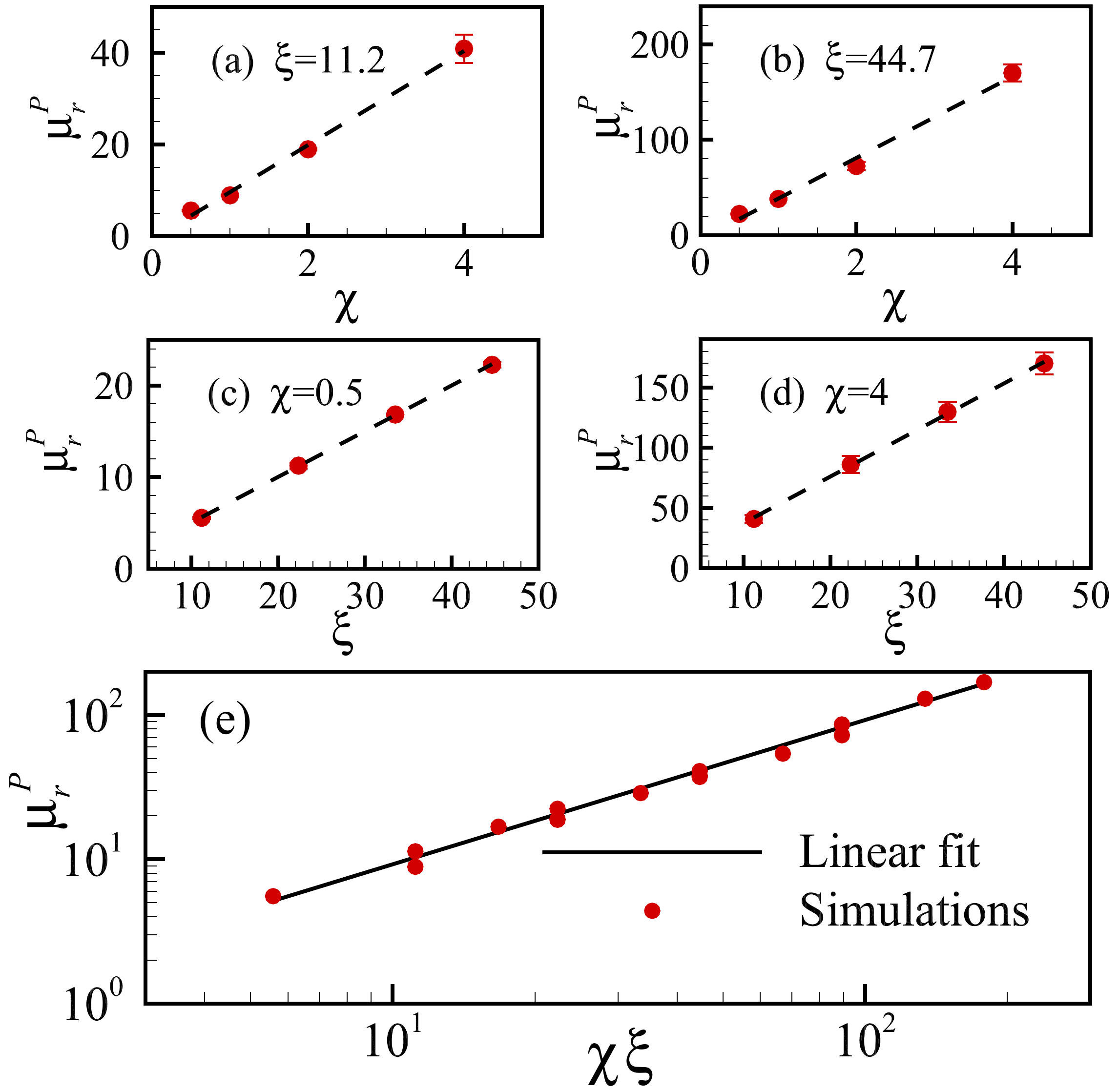,width=0.45\textwidth,clip=}}\hspace{0.0cm}  
	}
	\caption{Relative effective viscosity at the plug state, $\mu_r^P$. Panels (a)-(d):  $\mu_r^P$ as a function of the viscosity ratio (panels (a),(b)) and of the confinement parameter (panels (c),(d)), respectively. The dashed lines represent linear scalings. Panel (e): $\mu_r^P$ as a function of the product of viscosity ratio and confinement parameter, $\chi\xi$.  The solid line 
		represents the linear relation $\mu_r^P \sim \chi\xi$ in Eq. (\ref{e3}).}
	\label{Fig4}
\end{figure}

We then move to address the confinement and viscosity ratio dependence of the rheological response
more quantitatively, we focus next on the value of $\mu_r$ at the first plateau,
denoted as $\mu_r^P$. 
In  Fig. \ref{Fig4}(a) we plot $\mu_r^P$, averaged over the low $\Delta \hat{p}$ plateau, 
as a function of the viscosity ratio $\chi$, finding a linear scaling, $\mu _r^P \propto \chi $,
highlighted by the dashed line. This growth of the $\mu_r^P$ with $\chi$ is a
consequence of the lower deformability of droplets, as their intrinsic viscosity exceeds that of
the continuous phase. The linear scaling, which is less obvious and to which
we will return shortly, is preserved also for larger system sizes (Fig. \ref{Fig4}(b)),
albeit with a larger prefactor.
Analogous plots, showing the effective plug viscosity as a function of the confinement
parameter $\xi$, for two different viscosity ratios, are reported in Figs.
\ref{Fig4}(c-d).
In either case, $\mu_r^P$ increases linearly with $\xi$, $\mu _r^P \propto \xi $, with a
$\chi$-dependent prefactor.

We now proceed to rationalize the results on the plug viscosity 
under the light of simple phenomenological arguments grounded on mechanical 
considerations. Let us first notice that, since in the plug state the whole material moves as a solid, the mass
flux is $Q_P \approx U H$, where $U$ is the travelling speed shared by all droplet layers and 
whose value is dictated by the sliding velocity of the droplets on the walls, 
that we need to evaluate.
To this purpose, we recall that, when a pressure gradient $\Delta p/L$ is applied to a continuum system limited by solid walls 
(at a distance $H$ far apart), the force balance provides a wall stress $\sigma_w = \Delta p H/(2L)$.
A droplet in contact with the wall will be, then, subjected to a force $F \propto \Delta pHR/(2L)$, delivering 
a power input $\mathcal{P}_{\mbox{\tiny{in}}} \propto \Delta pHRU/(2L)$;
such input has to be balanced 
by the total dissipated power $\mathcal{P}_{\mbox{\tiny{diss}}}$ inside the droplet, which can be estimated as
the product of the viscous dissipation rate per unit volume $\varepsilon_{\mu_D} \approx \mu_D (U/R)^2$ and 
the droplet ``volume'' , $\propto R^2$ i.e. \cite{Podgorski,Kim}: $
\mathcal{P}_{\mbox{\tiny{diss}}} \sim \mu_D (U/R)^2 R^2 = \mu_D U^2. 
$ Although this estimate and the one for the force acting on a droplet are based on a 2D system, it is easy to notice that the scaling relation for $U$ remains the same
in 3D. Solving the balance equation $\mathcal{P}_{\mbox{\tiny{in}}} = \mathcal{P}_{\mbox{\tiny{diss}}}$ for $U$
yields: 
\begin{equation}
U\sim[\Delta p/(2L)]HR/{\mu _D},
\end{equation} 
delivering for the mass flux, 
\begin{equation}
{Q_P} \approx UH\sim[\Delta p/(2L)]R{H^2}/{\mu _D}.
\end{equation}
Since the mass flux in pure continuous phase Poiseuille flow
is ${Q_0} = (2H/3){U_0} = \Delta p{H^3}/(12L{\mu_C})$, we readily obtain for the relative effective plug viscosity $\mu_r^P$:
\begin{equation}\label{e3}
\mu_r^P \sim \frac{{{Q_0}}}{{{Q_P}}} \sim \frac{{{\mu_D}}}{{{\mu_C}}} \frac{H}{R} = \chi \xi.
\end{equation}
Eq. (\ref{e3}) informs us that $\mu_r^P$ grows linearly with both the viscosity ratio and the confinement parameter,
in agreement with the numerical results shown so far.

To check the validity of our prediction in Eq. (\ref{e3}), we plot in Fig. \ref{Fig4}(e) the values of $\mu_r^P$
as a function of the product $\chi\xi$, for all combinations of $(\chi,\xi)$ considered in this work, 
which span the ranges $\chi \in [0.5;4]$ and $\xi \in [11.2;44.7]$.  
We observe that all points tend to collapse, over a comparatively wide range of parameter values, 
onto a unique master curve which is well fitted by a single linear relation $\mu_r^P = A \xi\chi$,
in agreement with Eq. (\ref{e3}), with a prefactor $A \approx 0.92$.
\label{key}
\subsection{Effect of polydispersity}
\begin{figure}[!htbp]
	\center {
		\includegraphics[scale=0.43]{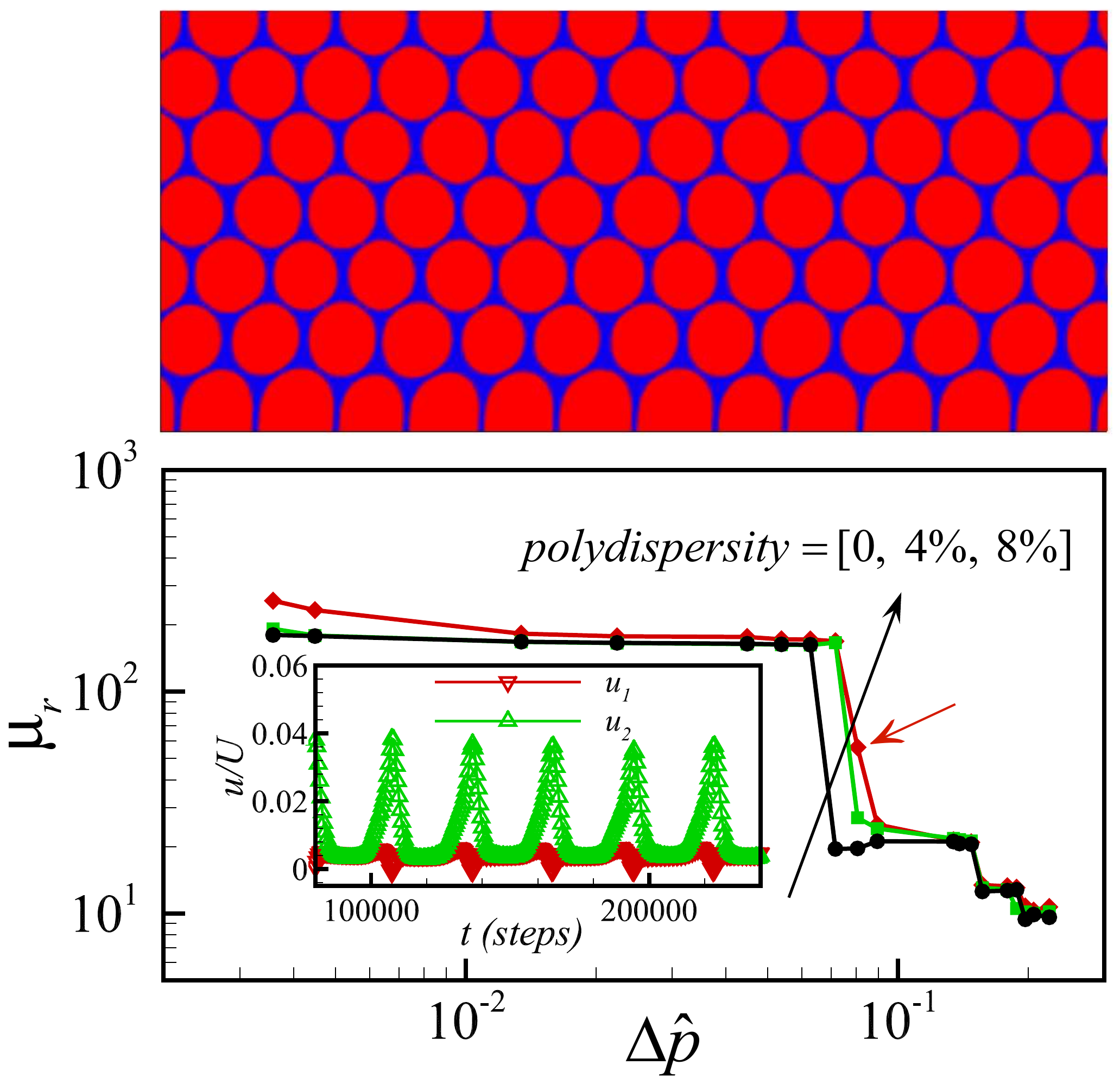}\hspace{0.0cm}  
	}
	\caption{Top panel: a snapshot of the $8\%$ polydispersity case with $\xi = 44.7$ and $\chi=4$.
			Bottom panel: relative effective viscosity, $\mu_r$, as a function of the rescaled pressure difference, $\Delta \hat{p}$, for three different volume polydispersities. A metastable state for the $8\%$ polydispersity case is marked by the red arrow. Inset: time evolution of the first two-layer droplets' velocities in the metastable state.}
	\label{fig5}
\end{figure}

Although the monodisperse emulsions have been used in many microfluidic systems \cite{Gai,vecchiolla2019,ISI:000459588200020}, the realization of perfectly monodisperse emulsions are essentially impossible in experimental systems.  Nevertheless, using the advanced experimental techniques \cite{ISI:000180449100018,ISI:000342866400008}, monodisperse droplets can be generated with very small volume polydispersity, e.g., less than $3\%$ \cite{ISI:000459588200020,gai2019timescale}. To link the present work with future experiments, it is necessary to discuss the effect of small volume polydispersity on the discrete fluidization process. To this end, we initialize the monodisperse emulsions, with some perturbations for each droplet's density ($ \rho  = {\rho _0} + \delta \rho $). As a result, a emulsion system with finite volume polydispersity can be produced in the steady state, and a snapshot for the case with $8\%$ polydispersity is shown in the top panel of Fig. \ref{fig5}. 

We then perform the same pressure driven simulations as before and obtain the relative effect viscosity curves for the produced polydisperse emulsions. We find that, for small and experimentally achievable volume polydispersity ($4\%$), the presented phenomenology is  well preserved, in the sense that we still observe a steep change in the relative effective viscosity around a characteristic pressure difference, but the ``plug''-``fluidized'' transition is gradually smoothed out with the increase of polydispersity (see Fig. \ref{fig5}). For example, we can see a metastable state for the $8\%$ polydispersity case (marked by the red arrow in Fig. \ref{fig5}), in the sense that the first and second layer droplets coincide and bifurcate periodically (in the inset of the bottom panel) and as a result the relative effective viscosity at the specified pressure gradient switches between the ``plug'' and ``fluidized'' state intermittently.

\section{Conclusions}\label{sec.4}
In summary, monodisperse crystalline dense emulsions in a pressure-driven microchannel flow have
been investigated numerically, using a new mesoscopic lattice kinetic model. Our study unveils the {discrete}
nature of the relative effective viscosity of this kind of soft materials under confinement and rationalizes their discrete 
fluidizations. We showed, for the first time, 
that such phenomenology is preserved in large (low confined) systems and even at changing the viscosity ratio 
between the two phases. We also provide physical interpretation of the phenomenon that the discrete fluidization is ascribed to local yielding evens.
Furthermore, we proposed a theoretical argument for the scaling relation of the relative effective viscosity of the ``plug'' state 
with confinement and viscosity ratio, and verified its validity over a wide parameter range.  

We would like to stress that our results might be of more general applicability. We expect that our model of SGMs, within suitable constraints on the imposed forcing, can capture  the physics of a broad family of monodisperse suspension of deformable particles (droplets and bubbles, as in foams, vesicles, etc) densely packed and confined in a microchannel.

Due to the simplicity of the configuration, it would be completely realistic to recreate the discrete fluidization in the lab, via suspensions of soft and non-coalescing droplets. In the future, it would also be informative to study the fluidization of dense emulsions or foams in more realistic  geometries \cite{gai2019timescale,vecchiolla2019}.

\section*{Conflicts of interest}
There are no conflicts to declare.
\section*{Acknowledgements}
The research leading to these results has received
funding from the MOST National Key Research and Development Programme (Project No. 2016YFB0600805) and the European Research Council under the European Union Horizon 2020 Framework Programme (No. FP/2014-2020)/ERC Grant Agreement
No. 739964 (COPMAT). Supercomputing time on ARCHER is provided by the UK Consortium on Mesoscale Engineering Sciences (UKCOMES)under the UK Engineering and Physical Sciences Research Council Grant No. EP/R029598/1.






\scriptsize{
\bibliography{GMRT} 
\bibliographystyle{rsc} } 

\end{document}